# A Systematic Mapping Study on Requirements Engineering in Software Ecosystems

Aparna Vegendla (Norwegian University of Science and Technology (NTNU), Trondheim, Norway),
Anh Nguyen Duc (Norwegian University of Science and Technology (NTNU), Trondheim, Norway),
Shang Gao (Örebro University, Örebro, Sweden) and Guttorm Sindre (Norwegian University of Science and Technology (NTNU), Trondheim, Norway)

**ABSTRACT**

*Software ecosystems (SECOs) and open innovation processes have been claimed as a way forward for the software industry. A proper understanding of requirements is as important for these IT-systems as for more traditional ones. This paper presents a mapping study on the issues of requirements engineering and quality aspects in SECOs and analyzes emerging ideas. Our findings indicate that among the various phases or subtasks of requirements engineering, most of the SECO specific research has been accomplished on elicitation, analysis, and modeling. On the other hand, requirements selection, prioritization, verification, and traceability has attracted few published studies. Among the various quality attributes, most of the SECOs research has been performed on security, performance and testability. On the other hand, reliability, safety, maintainability, transparency, usability attracted few published studies. The paper provides a review of the academic literature about SECO-related requirements engineering activities, modeling approaches, and quality attributes, positions the source publications in a taxonomy of issues and identifies gaps where there has been little research.*

## INTRODUCTION

The rapid pace of technological changes and the competitive race for quick product release are driving many companies to look for new ways to deliver software. Software product lines (SPLs) are one step towards making software development more efficient (Bosch & Bosch-Sijtsema, 2010). In SPL, a set of business units in an organization could develop the products through collaboration by sharing a common technological platform, and by reusing much of the software between different versions and variants of the product. Over the past decade, companies have been transitioning their SPLs to software ecosystems (SECOs) to open their platforms for external software providers (Bosch, 2009). The goal is to rapidly develop new capabilities and foster innovations unforeseeable by the platform's original designers (S. Jansen, and Michael A. Cusumano, 2013). The SECOs are multi-disciplinary systems inspired from business and natural ecosystems. Manikas and Hansen define software ecosystem as *"the interaction of a set of actors on top of a common technological platform that results in a number of software solutions or services"* (Manikas & Hansen, 2013)(p.1297). For example, Google controls the Android platform while external developers can build applications that are distributed to Android users via the Google Play store. Thus, Google has collaborated with external developers to build functionality in the form of available applications. In contrast to the software development in an individual organization, SECO includes the software development by several organizations through collaboration and competition (Bosch-Sijtsema & Bosch, 2015). For instance, Microsoft made the PowerShell tool built on Microsoft .NET as an open source product to keep the developers interested in the Windows platform while Google released Cloud Tools for PowerShell to make Google's cloud more attractive to .NET developers. Either way, both Google and Microsoft co-create value through collaboration and competition.

Despite the perceived advantages of SECOs, transitioning to SECOs may have challenges with communication barriers between parties due to the dispersion of SECO members. On the other hand, providing the open platform to external actors raises the conflicts of interest when negotiating requirements. One of the main issues is inconsistency and variability in stakeholders' requirements. Requirements engineering (RE) is essential for SECO's to involve stakeholders early in the development to understand requirements and use cases. The impact of changes can be analyzed and documented through a model of the system (Hull, Jackson, & Dick, 2011) during the early stage of RE. Modeling can aid the stakeholders of a SECO to make sustainable relations among the actors when they negotiate their common interests (i.e. requirements) for the software. The obvious question to ask is whether the RE process used for traditional systems can cope with the context of SECO's? How can the RE process best be conducted when developing multi-organizational, socio-technical systems like SECOs? Can adaptation of existing traditional approaches used in designing of technical system aids modeling of SECOs or does SECO require a new approach? Our research questions are: ***RQ1***: *What RE activities have been studied specifically in relation to software ecosystems earlier?* RE activities address, how are the requirements elicited, analyzed, documented, validated, and traced. ***RQ2:*** *How are non-functional requirements considered in the context of SECOs?* A number of challenges in SECO's have been discussed in the literature (Serebrenik & Mens, 2015), we focus on challenges specific to non-functional requirements in SECOs.

The remainder of the paper is structured as follows. Section 2 provides the background of SECOs, RE, and quality in SECO. In Section 3, we describe the research method used for conducting a literature review for the paper. Section 4 analyzes the results. Section 5 provides the discussion of research gaps identified in the existing literature on RE in SECOs follows with conclusion in section 6.

## THEORETICAL BACKGROUND

### Software Ecosystems

Some previous studies (Barbosa & Alves, 2013; Bosch, 2009; Jansen, Finkelstein, & Brinkkemper, 2009) provided an overview of the key concepts and implications of adopting a SECO. Multiple definitions of a SECO exist (Jansen, Brinkkemper, & Finkelstein, 2009). This paper will use the one by Jansen et al. (S. Jansen et al., 2009), as "*a set of actors functioning as a unit and interacting with a shared market for software and services, together with relationships among them. These relationships are frequently underpinned by a common technological platform or market and operates through the exchange of information, resources and artifacts*." During the last decade, there has been a lot of research projects focusing on the concepts of roles and relationships involved in SECOs. Manikas and Hansen (Manikas & Hansen, 2013) performed a more detailed study on SECOs through a systematic literature review. Their systematic literature study was later extended by Manikas (Manikas, 2016) to provide an updated overview of the SECOs. Apart from architecture aspect of SECOs (Bosch, 2009, 2010; Bosch & Bosch-Sijtsema, 2010; García-Holgado & García-Peñalvo, 2016; Hartmann & Bosch, 2014; Rajeshwar, 2017; Walt Scacchi & Alspaugh, 2012b), recent studies investigate business and actors perspectives of SECOs (Fricker, 2010; Valença, Carina, Virgínia, Slinger, & Sjaak, 2014; Van den Berk, Jansen, & Luinenburg, 2010), i.e. software supply networks (Joey van Angeren, Blijleven, & Jansen, 2011)**,** and collaboration patterns among SECO actors (Cataldo & Herbsleb, 2010; Knauss, Damian, Knauss, & Borici, 2014). There are also studies of geographical distribution and management of engineering practices (Goeminne, 2014; R. P. d Santos & Werner, 2012; W. Scacchi, 2007; Teixeira & Lin, 2014). While considering the requirement engineering perspective of SECO, we emphasized the coordination of requirement engineering activities in relation to multiple stakeholders. According to Hanssen (Hanssen & Dybå, 2012), the three roles the organization may have in the ecosystem are:

- *Keystone organization* which leads the development.
- *End-users* of the central technology, needing it to carry out their business.
- *Third-party organizations,* using the central technology as a platform for developing related solutions and services (Hanssen & Dybå, 2012).

### Requirements Engineering

Requirements engineering (RE) is a subfield of Software Engineering, dealing particularly with how to find and specify requirements for software and software-intensive systems. Loucopoulos and Karakostas (Loucopoulos & Karakostas, 1995) define the RE as "*systematic process of developing requirements through an iterative co-operative process of analyzing the problem, documenting the resulting observations in a variety of representation formats and checking the accuracy of the understanding gained*." RE is a crucial process of the system development because errors made in the requirements specification will be the more costly to correct the longer they stay in the project (i.e., through design, testing, release). The processes used for RE vary depending on the type of system being developed. However, some activities common to all processes are:

- **Requirements elicitation** encompasses the extraction of requirements from stakeholders, reaching the understanding of the stakeholders' needs, understanding of the domain within which the system is embedded, and understanding of the constraints (i.e. non-functional requirements) that can be placed on the system (Loucopoulos & Karakostas, 1995). This activity includes the task of developing mental models that describe the domain.
- **Requirements analysis** determines whether the requirements are clear, complete, and unambiguous. In a case of conflicting requirements during the analysis, negotiation sub-activity provides the agreement among the stakeholders. This activity runs in parallel with elicitation, specification and validation activities.
- **Requirements specification (or documentation)** analyzes the acquired knowledge in the elicitation and transform the informal requirements to formal requirements, and model the requirements into a formal model called requirements specification model. The requirements specification model cannot be developed in a linear fashion, but a cyclic approach gradually improves the model (Loucopoulos & Karakostas, 1995). This activity controls both elicitation and validation.
- **Requirements validation** ensures that the produced formal requirements model satisfies the users' needs. Validation is applied on not only the formal model but also to the informal requirements from elicitation (Loucopoulos & Karakostas, 1995).
- **Requirements management** (Hull et al., 2011) includes the task of the management of requirement changes after the requirements elicitation and specification. Requirements analysis and management should be conducted in parallel to handle the conflicts and ambiguity in the requirements.

**Product Quality in SECO**

Whereas software ecosystem is a relatively new phenomenon, research about product quality has been intensive in Software Engineering (SE) literature. There are many different definitions of quality. One of the eldest definitions of quality states that "*there are two common aspects of quality: one of them has to do with the consideration of the quality of a thing as an objective reality independent of the existence of man. The other has to do with what we think, feel or sense as a result of the objective reality. In other words, there is a subjective side of quality*" (Shewhart, 1930). SE research and practitioners acknowledged the differentiation of software functional quality and software structure quality. Software functional quality is about how well it complies with or conforms to a given design, based on functional requirements or specifications. Software structural quality refers to how it meets non-functional requirements that support the delivery of the functional requirements, such as reliability or maintainability, the degree to which the software was produced correctly (Pressman, 2005). Both these definitions concern primarily how the product is performing during its operational use, and this is also the emphasis of this paper.

The structure, classification, and terminology of attributes and metrics of software quality are captured in various quality models. McCall's quality model proposed three major perspectives for defining and identifying the quality of a software product: product revision (ability to undergo changes), product transition (adaptability to new environments) and product operations (its operation characteristics) (McCall,

Richards, & Walters, 1977). Other models might also be mentioned, such as those by Boehm's (Boehm et al., 1978) and Dromey's (Dromey, 1995). Based on the McCall and Boehm models, ISO released the ISO 9126: Software Product Evaluation: Quality Characteristics and Guidelines for their Use-standard (Standardization & Commission, 2001). ISO 9126 proposes a standard which species six quality attributes, namely functionality, reliability, efficiency, maintainability, portability, and usability. In addition, a lot of SECO research investigates the software architecture aspect (Bosch, 2010; García-Holgado & García-Peñalvo, 2016; Hartmann & Bosch, 2014; Rajeshwar, 2017), where it is a common practice to discuss the "quality attributes" of an architecture. We also differentiate the product quality as a characteristic of software platform or artifact created by the SECOs and the ecosystem quality, i.e. *ecosystem health*, which described the organizational perspective of SECOs.

## SYSTEMATIC MAPPING STUDY

### Methodology

We planned, conducted, and reported review findings for the papers published from 2009 to 2017 by following the SLR process suggested by Kitchenham et al. (Kitchenham, Budgen, & Pearl Brereton, 2011). Our study is characterized as a systematic mapping study, as we did not aim at performing any meta-analysis of primary studies (Hartmann & Bosch, 2014). The systematic mapping study classifies the relevant literature in that particular domain and aggregates studies on defined categories e.g. author's names, authors affiliations, publication source, publication type, publication date, etc. (Kitchenham et al., 2011). Moreover, we also extracted the state-of-art research on RE activities and non-functional requirements in SECO. The search process is conducted in five steps, as the way we had performed SLR in our previous studies (Nguyen-Duc, 2017; Nguyen-Duc, Cruzes, & Conradi, 2015):

- **Step 1: Seeding key research**. We had already a list of paper about requirement modeling and security issues in SECOs as our seed papers. From the list, we expand the scope to cover the current state-of-the-art about Requirement Engineering in SECO context.
- **Step 2: Defining search protocol.** The search strings were experimented and derived from our RQs. As our objective was to identify RE activities, and especially non-functional requirement activities, we used the synonyms them in the search string. We considered alternative terms for "software ecosystem", such as "software supply chain" or "software supply networks". However, the additional terms did not introduce any more relevant studies, which is also observed in another SLR (Axelsson & Skoglund, 2016). Eventually, the final search string was used as below:

    *("software ecosystem") AND ((requirement OR specif\* OR model\* OR verif\* OR valid\* OR elicit\* OR analy\* OR negotiate OR document OR manag\*) OR ("non-functional requirement" or "software quality" or security or safety or usability or maintainability or testability or performance or reliability))*

    The search string was a pilot and tested in Scopus index database to validate search's coverage. The inclusion and exclusion criteria were defined as in Table 1. The limitation was that we only searched for papers written in English and papers that full contents are not accessible.

*Table 1. Inclusion and exclusion criteria*

| Inclusion criteria | Exclusion criteria |
|---|---|
| • Studies within Software Engineering, Computer Science, Information System or sub domains <br> • Studies investigating either RE activities, RE related challenges, modeling techniques in SECO context <br> • Studies investigating non-functional requirements, i.e. security in SECO context <br> • Both industrial reports, empirical studies, theoretical papers | • Position or short papers <br> • Studies about technical aspects, i.e. algorithm, network. <br> • Studies not written in English |

- **Step 3: Conducting a systematic search**. We tailored the search string to fit to the four different search databases, Scopus, IEEE Xplore Digital Library, ACM Digital Library, and Elsevier Science Direct. The actual systematic search was performed, using search protocol developed from Step 2. We searched in the metadata fields Title, Abstract, and Keywords when available. There were differences in search possibilities among the digital libraries and thus the search queries varied slightly.

The result of this step was a set of 1676 papers. After a unique set of primary studies was retrieved, the inclusion and exclusion criteria were used to filter out the non-relevant papers. The paper selection was done by two filterings:
  - Filter 1: The returned papers were scanned by reading the title, abstract, keywords, and conclusion, to eliminate irrelevant papers.
  - Filter 2: Look through the full text of the papers from Filter 1 with selection criteria.

Filter 1 was conducted mainly by the first author. Filter 2 was done separately by the first author and the second author. A subset of the selected papers was used to check the consistency among us. During the selection, we also consider duplicate papers and repeated studies. In the case of duplicate papers, we retain only paper accepted on a site with a higher rating. We excluded papers that do not explicitly investigate the mentioned topics. The search result was shown as in Table 1.

*Table 2. Number of selected studies per phase*

| Source | Returned | Filter1 | Filter2 |
|---|---|---|---|
| IEEE Xplore Digital Library | 208 | 18 | 7 |
| ACM Digital Library | 562 | 31 | 11 |
| Elsevier Science Direct | 497 | 24 | 19 |
| Scopus | 509 | 20 | 4 |
| Manual search | | 4 | 3 |
| Total | 1676 | 97 | 44 |

- **Step 4: Additional manual search**. A manual search was performed to retrieve more relevant papers and to search for grey literature (unpublished papers, white reports, master thesis, company documents that are not published as scientific work). The importance of such literature has been emphasized in multivocal literature review approach (Garousi et al., 2016).The reference list of selected papers was scanned in order to find more relevant papers. The manual search in related conference and journals was also conducted. In the end, the additional manual search in Google added some extra papers that were not indexed by Scopus, IEEE Explore or ACM Digital library or not published. In the end of this step, we identified four extra relevant papers
- **Step 5: Quality assessment**. Paper quality assessment was conducted to remove low quality papers. In this step, we removed papers that were identical in the data set. The quality assessment form was reused from our previous study (Nguyen-Duc et al., 2015).

Table 2 presents the number of papers retrieved and selected in each phase of the study. The selected 44 papers can be seen as the comprehensive collection of publications related to the research in RE in software ecosystems. The search process is summarized as in Figure 1:

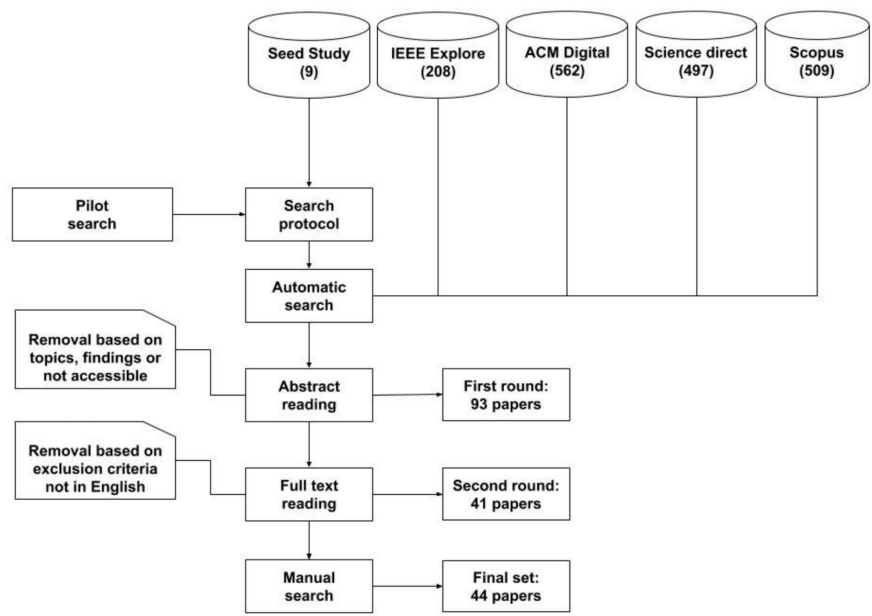

*Figure 1. Search Process*

## Threats to validity

One common threat to systematic literature reviews is not to discover all relevant studies. To reduce this threat, we defined the search strings to retrieve as many documents as possible which were related to the research topic. By extending the search scope, we achieved 1676 documents from the search engines, which reflected a very low search precision. In addition, the other search done by using the Google Scholar search engine and forward, backward scanning through the relevant studies helped to further decrease the probability of missing relevant studies. We believe that the group of missing papers is small enough to not influence the findings from our review.

A relevant study may be misclassified into a removal group during the selection process and vice versa. To reduce the bias in the selection of papers, we defined a review protocol with clear inclusion and exclusion criteria for each selection phases. We also adopted a well-prepared quality assessment checklist to judge the paper's quality (Nguyen-Duc et al., 2015). The selection and extraction were done by at least two authors together, which helps to reduce personal bias.

# RESULTS
## Demographic of primary studies

44 primary study papers used for mapping study are presented per year from 2009 to 2017 in Table 4. The first included paper was published in 2009. The mapping study revealed that at least three papers were focusing on RE in SECOs each year from 2009 and this is a potential sign that the research on RE of SECOs is gaining increasing attention among researchers and practitioners in software ecosystems. The peak publication period is from 2013 to 2015 (22 papers, around 50%).

Table 3 shows the publication channels of the selected studies. The selected studies have been published in 5 journals ((Axelsson & Skoglund, 2016; Christensen, Hansen, Kyng, & Manikas, 2014; Eklund & Bosch, 2014; Fernandez, Yoshioka, Washizaki, & Syed, 2016; Gherbi, Charpentier, & Couture, 2011; Leite & Cappelli, 2010; Mhamdia, 2013; Walt Scacchi & Alspaugh, 2012b)), 16 conferences ((Bezzi, Damiani, Paraboschi, & Plate, 2013; Bosch, 2010; Campbell & Ahmed, 2010; Claes, Mens, & Grosjean, 2014; Dantas, Marinho, Costa, & Andrade, 2009; Fahl et al., 2014; Fernandez, Yoshioka, & Washizaki, 2015; Fricker, 2010; Goldbach & Benlian, 2015; H. Sadi & Yu, 2015; Handoyo, Jansen, & Brinkkemper, 2013; Hansen & Zhang, 2013; S. Jansen et al., 2009; Knauss et al., 2014; Knauss, Yussuf, Blincoe, Damian, & Knauss, 2016; Pettersson, Svensson, Gil, Andersson, & Milrad, 2010; Sadi & Yu, 2017; Rodrigo Pereira dos Santos, 2014; R. P. d Santos & Werner, 2012; Walt Scacchi & Alspaugh, 2012a; K.-B. Schultis, Elsner, & Lohmann, 2014; Valenca, 2013; Valença et al., 2014; Van den Berk et al., 2010)), 7 workshops ((J. van Angeren, Kabbedijk, Jansen, & Popp, 2011; Boucharas, Jansen, & Brinkkemper, 2009; Fricker, 2009; S. Jansen, 2013; Slinger Jansen et al., 2009; Jansen, Handoyo, & Alves, 2015; K. B. Schultis, Elsner, & Lohmann, 2013; Serebrenik & Mens, 2015; Soltani & Knauss, 2015; Walden, Doyle, Lenhof, Murray, & Plunkett, 2010; Yu & Deng, 2011)), and one book chapter (Jansen & van Capelleveen, 2013). These studies are from three main fields, namely RE, SE, and IS.

The publication channels contributing more than two studies are Requirements engineering conference and European conference on software architecture, Journal of systems and software, International workshop on software ecosystems. The authors contributing more than two studies are Slinger Jansen (8 papers, 18 %), Eric Yu (3 papers, 6%). This author list is represented based on the result of our mapping study.

*Table 3. The publication channels of primary studies*

| Conference | Paper Count | Journal | Paper Count | Workshop | Paper Count |
|---|---|---|---|---|---|
| ICSE | 2 | JSS | 3 | ECSAW | 2 |
| RE | 4 | IST | 1 | IWSECO | 4 |
| ICIS | 1 | IJPPM | 1 | IWOCE | 1 |
| ESPRE | 1 | Future internet | 1 | WEA | 1 |
| ECSA | 5 | CrossTalk | 1 | IWSMM | 1 |
| REFSQ | 2 | BISE | 1 | EmpireRE | 1 |
| ICSOC | 1 | | | PLEASE | 1 |
| EMMSAD | 1 | | | | |
| SIGSAC | 1 | | | | |
| ICSOB | 1 | | | | |
| ISCIS | 1 | | | | |

| | | | |
|---|---|---|---|
| ICGSEW | 1 | | |
| ICOSS | 1 | | |
| SIGSOFT | 1 | | |
| CSP | 1 | | |
| CSMR-WCRE | 1 | | |
| | 24 | 8 | 11 |

*Table 4. The primary studies published per year*

| Year | Papers | Total |
|---|---|---|
| 2009 | [9, 10, 59, 65, 70] | 5 |
| 2010 | [13, 17, 18, 45, 49, 61, 71] | 7 |
| 2011 | [40, 66, 67] | 3 |
| 2012 | [12, 24, 53] | 3 |
| 2013 | [42, 52, 55, 60, 68, 69, 72] | 7 |
| 2014 | [19, 22, 41, 43, 46, 54, 56-58] | 9 |
| 2015 | [7, 47, 48, 51, 63, 64] | 6 |
| 2016 | [38, 44, 62] | 3 |
| 2017 | [50] | 1 |
| Total | | 44 |

*Table 5. Research topics across RE activities in SECOs*

| Activity | Topics | Papers |
|---|---|---|
| Elicitation | Goal modeling | [67] |
| | Reference model | [18, 49] |
| | Non-functional requirements | [38, 53] |
| | Identifying stakeholders' roles | [10, 17, 52, 65, 66] |
| | Identifying relationship | [10, 54, 66] |
| | Policies | [12] |
| Analysis | Requirements communication or Negotiation | [17, 22, 60, 65] |
| | Conflict management | [17, 19, 54] |
| | Conflict analysis | [43, 65, 69] |
| | Requirements prioritization | [9, 22] |
| | Requirements selection | [19] |
| Specification | Notation semantics | [70] |
| | Modeling approaches | [18, 24, 46, 49, 50, 67, 70] |
| Validation | Model formalism | [70] |
| | Requirements verification, validation and testing | [64] |
| Management | Global RE | [24] |
| | Management practices | [62] |

## RQ1: What RE activities have been studied specifically in relation to software ecosystems earlier?

Table 5 presents results from mapping study on RE activities. The table indicates what topics have been covered in the SECO-related research.

**Requirements Elicitation.** Elicitation involves identifying SECO members' requirements according to their business goals. As software vendors have trouble distinguishing the particular software ecosystem they are active in, and are having trouble using SECO advantages for their strategic planning (Boucharas et al., 2009), the requirements elicitation in a software ecosystem context can be considered a challenging issue. The openness of the platform permits everyone to provide the requirements without role identification. Several studies (J. van Angeren et al., 2011; Fricker, 2009, 2010; Handoyo et al., 2013; Slinger Jansen et al., 2009) discuss the techniques for identifying stakeholders' roles. To represent the goals and strategic interests of SECO actors, Yu and Deng (Yu & Deng, 2011) proposed a modeling approach for SECOs based on the i* modeling language. Several studies (Pettersson et al., 2010; Van den Berk et al., 2010) propose reference models to model the relationship between SECO members in domain specific software ecosystems.

*Table 6. Results type per solution studies*

| Result | Papers |
|---|---|
| Technique | [67] |
| Procedure | [49] |
| Conceptual architecture | [46] |
| Reference model | [18] [49] |
| Notation | [70] |

The RE activity in SECOs involves an understanding of software component licenses that are part of the composed system after the system integration. The software licenses of the components both aid and constrain the SECO evolution. Scacchi and Alspaugh (Walt Scacchi & Alspaugh, 2012b) described an example of the open architecture software system which provides the niche developers with guidance for identifying evolutionary paths of the system and architecture.

The non-functional requirements pertaining to the SECOs are least addressed in the literature. Our search results indicate only three studies (Axelsson & Skoglund, 2016; Walt Scacchi & Alspaugh, 2012a) that discussed the non-functional requirements for SECOs. Scacchi and Alspaugh (Walt Scacchi & Alspaugh, 2012a) discussed an approach to specify and analyze security requirements conflicts in open source components context in SECOs through security license. Axelsson and Skoglund (Axelsson & Skoglund, 2016) provided the mapping study on quality assurance. There are no publications on requirements elicitation tools and techniques particularly for SECOs.

**Requirements Analysis.** Requirements communication in SECOs is challenging when stakeholders in SECOs need to communicate from dispersed locations. Several studies (Fricker, 2009, 2010; Valenca, 2013) discuss the requirements negotiation strategies in SECOs. The negotiation and network theory were among the theories adopted for the analysis and design of the requirements among SECOs members (Fricker, 2009, 2010). The network theory was used to evaluate the strengths and weakness of the stakeholder's network, and negotiation theory was utilized for the decision-making knowledge.

A modeling approach by Fricker (Fricker, 2009) proposed the Requirements Communication Network (RCN) which describes the structure of SECOs through the notation, modeling language, and framework based on negotiation and network theory. He later proposed the concept called requirement value chains (Fricker, 2010). This concept is particularly useful to analyze the requirements and agree on a set of requirements. This kind of approach especially improve the communication bottlenecks among SECO members.

Valenca (Valenca, 2013) proposed a requirement negotiation model considering the social perspective for addressing the negotiation activities with the platform management in SECOs. Valenca claimed that the recent research in SECOs is not considering the integration of ecosystems' social dimensions to a business view during the negotiations, and the research is generally concerned with the challenges of achieving the negotiated requirements. Valenca et al. (Valença et al., 2014) later investigated the impact of tightening relationships on software product management with a focus on RE practices.

During the requirements analysis, the conflict management (Fricker, 2010; K.-B. Schultis et al., 2014; Valença et al., 2014) and conflict analysis (Christensen et al., 2014; Fricker, 2009; K. B. Schultis et al., 2013) were addressed. While there is only a concise information is addressed on requirement prioritization (S. Jansen et al., 2009) and selection (Valença et al., 2014) in the ecosystems.

**Requirements Specification.** Several studies discussed on SECOs documentation with SECOs model using different views (Campbell & Ahmed, 2010; R. P. d Santos & Werner, 2012), and approaches for the modeling. Campbell and Ahmed (Campbell & Ahmed, 2010) proposed the structure of SECOs through the three-dimensional view. The structure was further extended by Santos and Werner (R. P. d Santos & Werner, 2012) with the 3+1 framework named ReuseECOS with additional management and engineering (M&E) view. The M&E dimension facilitates the global software development (GSD) in SECOs.

The SECOs modeling is addressed by many researchers. Boucharas et al. (Boucharas et al., 2009) propose a software ecosystem modeling (SEM) approach enabling software vendors to communicate about relationships in the software supply network. The approach consists of two components: SSN (software supply network) and PDC (product deployment context) for communication. The paper describes that the SSN is specifically proposed for SECOs to deal with the actors and their relationships. Though the SEM approach provides the model formalism and notation semantics, it does not include a tool for modeling. Van den Berk et al. (Van den Berk et al., 2010) represent the SECOs key characteristics by a model called SECO Strategy Assessment Model (SECO-SAM). An approach to support the definition, modeling, and analysis of SECOs through software reuse concepts in SECOs was proposed (Rodrigo Pereira dos Santos, 2014). Santos presented the approach through the conceptual architecture. Sadi and Yu (H. Sadi & Yu, 2015) reviewed several published modeling techniques and further they proposed an approach (Sadi & Yu, 2017) to model and analyze the trade-offs between openness requirements and non-functional requirements. However, the lack of the universally accepted set of modeling methods is hampering the advancement of SECOs (Jansen et al., 2015).

**Validation and Verification.** Well-organized specification reviews are essential for a successful relationship between SECOs members. Soltani and Knauss (Soltani & Knauss, 2015) performed verification in the cross-organizational requirements engineering process through specification reviews on the automotive open system architecture (AUTOSAR) standard.

**Requirements Management.** As SECO members are globally distributed, requirement management is challenging. Requirement management activity comprises a number of activities related to managing requirements e.g. traceability and variability in requirements. The requirement management specific to traceability and variability of the requirements negotiated at the elicitation was not described in the collected papers.

### RQ2: How are non-functional requirements considered in the context of SECOs?
Quality assurance cannot be seen as an isolated activity, but rather as a set of activities and processes that take place during different phases of the software life-cycle (Axelsson & Skoglund, 2016). The high-level process areas, namely agreement processes, organizational project-enabling processes, project processes, and technical processes, with sub-processes defined for each of them. For example, the technical processes include activities such as requirements definition, system architectural design, implementation, integration, and operations. Quality assurance related activities can occur in any of these processes.

The quality of a software product can be measured using different quality attributes. In the paper, we explain some of the quality attributes of the software in the ecosystem, including Reliability, Maintainability, Safety, Security, Performance, Testability, and Usability. The research topics across software quality attributes in SECOs are represented in Table 7.

**Reliability.** The reliability concerns in ecosystems have been discussed by Bosch (Bosch, 2010). He described that the composition of platform and extensions developed on the top of the platform leads to security and reliability concerns as the malicious code spread from extensions to the platform. The Dependencies between unreliable components in Comprehensive R Archive Network (CRAN) ecosystem have been addressed in (Claes et al., 2014).

**Safety.** Many of the software ecosystems come from mobile applications in IT domain. The examples of the safety of software ecosystems have been studied based on embedded systems domain (Eklund & Bosch, 2014).

*Table 7. Research topics across software quality attributes in SECOs*

| Quality Attribute | Type | Topics | Papers |
|---|---|---|---|
| Reliability | In-operation | Reliability concerns<br>Dependencies between unreliable components in CRAN ecosystem | [13]<br>[56] |
| Maintainability | In-development | Dependency problems in CRAN software ecosystem | [56] |
| Safety | In-operation | Architecture for embedded open software ecosystems, Safety of automotive software | [41] |
| Security | In-operation | Certification and policy management<br>Security patterns<br>Software redundancy<br>Authentication, accountability, transparency | [55]<br>[44, 48]<br>[40]<br>[57] |
| Performance | In-development | Developer performance<br>Performance measurements of eclipse software ecosystem<br>Ecosystem healthiness: robustness, productivity, interoperability, stakeholder satisfaction and creativity | [47]<br>[58]<br><br>[42] |
| Testability | In-development | Software testing requirements for mobile applications<br>Static analysis (or Code reviews) of vulnerabilities in web applications and their plugins<br>Review and approval methods for platform extensions | [59]<br>[71]<br><br><br>[72] |
| Transparency | In-development<br>In-operation | Requirements engineering<br>confidentiality | [45]<br>[62] |
| Usability | In-operation | Software platform usability and extendibility | [68] |

**Security.** Security requirements are considered important parameters for defining, designing, and assessing the architecture of the system. The opening of software platform architecture can ease the extension of its functionality. However, the platform in SECOs experience increased serious security risks from the visibility of platform functionality to the third party applications. For instance, malicious code spread from externally developed third-party applications to the platform may cause the overall system to become disabled (Bosch, 2010). A good architecture can minimize the likelihood for malicious code to affect the whole system, but the defects can only be minimized to the extent that the security mechanisms are correctly implemented. For instance, the confidentiality and integrity of data can be ensured only when the security certification (Bezzi et al., 2013) is correctly used by external applications (Walt Scacchi & Alspaugh, 2012a).

Security and privacy can also be evaluated using the architectural models that are represented in the form of pattern diagrams including software patterns (Fernandez et al., 2015; Fernandez et al., 2016). Pattern models are especially useful for handling the security of complex systems like SECOs.

Another way is to have the critical systems implemented in multiple instances. Software redundancy (Gherbi et al., 2011) can improve identification of the vulnerability exploits and can handle the user requests when one system is targeted for a security attack.

**Performance.** Performance of the ecosystem is measured through ecosystem healthiness considering various dimensions, including robustness, productivity, interoperability, stakeholder satisfaction and creativity (Mhamdia, 2013). The combined dimensions enhance the quality of the decision making process. The analysis of performance characteristics of releases of Eclipse software ecosystems was performed by (Hansen & Zhang, 2013).

The performance of software ecosystem is also depends upon external developers intention to be a part of the ecosystem. Goldbach and Benlian (Goldbach & Benlian, 2015) performed a study to provide a deeper understanding of the positive effects of informal control modes i.e. self-control and clan control in software platform settings that reveals a mediating role of third-party developers intrinsic motivation on developers effort and intention to stay on a platform.

**Testability.** The common problem being observed in SECOs is that the vulnerabilities transferred from plug-ins or add-ons to the main application and vice versa. Software evaluation is ensured using a number of methods, including verification of compliance to policies, testing, code reviews etc.

Considering the issue with the vulnerability spread from platform and external developers, the study on the vulnerability analysis has been performed on 12 open source PHP web applications using the static analysis technique (Walden et al., 2010). The process of static analysis was conducted through code reviews on the source code of the applications.

**Transparency.** Transparency is an important non-functional requirement to maintain the security in ecosystems especially the confidentiality of the code and information shared in the ecosystem. "The more the transparency is, lesser the confidentiality becomes".

Open source code does not guarantee transparency (Leite & Cappelli, 2010). In SECOs transparency depends on how open code is provided for platform extension. For instance, only part of the code is available for use depending on guidelines to use the code.

Transparency has also been discussed in open commercial ecosystems related to confidentiality when maintaining the openness and transparency in the ecosystem (Knauss et al., 2016).

**Usability.** Currently, there is little knowledge available on how a successful and easy-to-adopt software platform architecture can be developed. Jansen (S. Jansen, 2013) provides insight into software platforms and platform usability for platform developers and for third-party developers (i.e., platform extenders). Furthermore, he studied the relation between platform adoptability by platform extenders and a platform's architecture.

The results of non-functional requirements are shown in Figure 2.

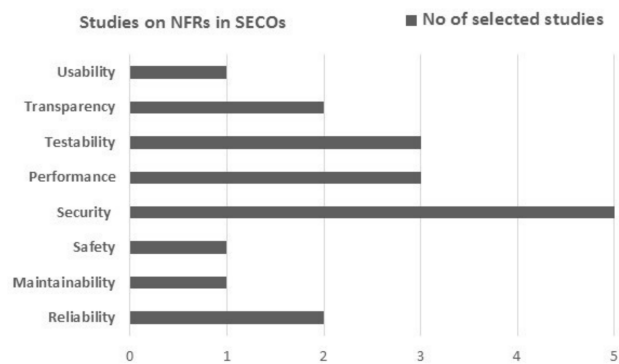

*Figure 2. Number of selected studies concerning Non-functional requirements*

## DISCUSSION

SECOs have attracted the attention of both academics and practitioners. This section discusses the results of this Systematic Mapping Study: an integrated and synthesized answer is provided in response to the RQs; contributions to the SECO field are highlighted. Concerning the answer for RQ1, we provided the different RE activities and challenges on RE in SECOs, as discussed below.

A large scale SECO require the platform to be available to external developers outside the organization which additionally requires new business models for negotiation between stakeholders (Bosch, 2010; S. Jansen et al., 2009), and awareness among developers for the evolution of SECOs (Lettner & Grünbacher, 2015). During the last decade, most of the research in SECOs has been done on architectural challenges (Hartmann & Bosch, 2014) and high-level views used to address different perspectives of the stakeholders. Nevertheless, high priority on architectural model and views (Campbell & Ahmed, 2010; R. P. d Santos & Werner, 2012) could not provide the design which ensures success of the organizations. Requirements analysis and management should be conducted in parallel to handle the conflicts and ambiguity in the goals. Our mapping study shows that from 2011, there is more progress on architectural design issues in SECOs that are explained through models and different dimensions. The success of RE depends on all activities in the RE process but not on the requirements architectural modeling alone. A systematic way of performing an activity is not possible without an RE method and guidelines to conduct RE process in SECOs.

Several studies propose reference models (Pettersson et al., 2010; Van den Berk et al., 2010), conceptual architectures (Rodrigo Pereira dos Santos, 2014), and dimensions, rather than requirements analysis approaches. We still included these in selected studies because the insight that these works offer into the structure and views of the SECOs can be used by a requirement modeling method to understand the stakeholder requirements from different dimensions and to perform RE activities for SECOs. The SEM approach of Boucharas et al. (Boucharas et al., 2009) is difficult to understand without technical expertise. There could be more understanding of the requirements when strategic models are developed through techniques like i* (Yu & Deng, 2011) for SECOs.  The lack of guidelines for the RE process in SECOs could additionally require more effort from stakeholders in negotiating the requirements for the complex systems being developed in SECOs. We found that the requirements elicitation, analysis, specification activities are most addressed in the literature. For requirements analysis, only one paper addressed requirements selection, and there are no papers published on the tools to analyze the requirements e.g., ontologies, risk management, and variable requirements analysis. For requirements modeling, one publication provided certain notations and semantics for modeling, but there are no papers on the patterns to specify the users' requirements and the tools to model the requirements. The requirements validation, and management activities are least addressed. Nevertheless, without validation and management of the requirements, the consistency and completeness of the requirements and models, and the changes in the requirements are not ensured. The traceability is a significant task to provide the changes in the requirements. Hence, more research should be required for the requirements verification and management.

Concerning the RQ2, we provided insights on non-functional requirements for the quality of the platform extensions in SECOs. Most of previous research tended to focus on the following types of non-functional requirements in the SECOs context: reliability, maintainability, safety, security, performance, testability, transparency, and usability. Security was the most addressed concern in the previous research on this from 2009 to 2017.

Non-functional requirements in the context of SECOS were often discussed with an association with architecture of software ecosystems. There are many stakeholders in the software ecosystems. It is also important to make collected non-functional requirements in the SECO context transparent to all involved stakeholders. Software architects prefer to use extended platform architecture to take advantages of extensions. The malicious code and virus spread from extensions can be prevented by using different security policies for extensions because they run their own processes (Jansen & van Capelleveen, 2013). This kind of loosely coupled architecture can increase quality of platform, platform extendibility and ecosystem performance.

According to the analysis results, we found that there are two types of quality concerns: In-operation, and in-development. In-operation quality concerns occur during the use, operation and maintenance of the software product as the outcome of the SECOs. Reliability, safety, security, transparency, and usability are the major concerns for the in-operation phase. In-development quality concerns occur during the development of the platform/products, which often relate to multiple actors in SECO. Maintainability, performance, testability, transparency, and usability are the major attributes related to the in-development phase. Some attributes appear in both forms, with different focus. For example, in-operation security concerns tend to be about user behavior while in-development security tends to focus on secure development processes and practices, and the behavior of developers. In-operation transparency implies concerns about confidentiality of user data, while in-development transparency implies concerns about requirements engineering practices, use of the platform/source code etc.

The study of quality in SECOs so far lacks the investigation of domain-specific needs. Mobile SECOs might have different quality focus than embedded SECOs. For instance, in embedded SECO, the focus is on network level security, while in mobile SECO, the focus is on software security.

**CONCLUSION**

Software ecosystems has emerged as an important research topic recently. This paper has provided a basis for synthesizing existing knowledge for RE in software ecosystems (SECOs) by a mapping study of publications addressing or relating to RE and non-functional requirements for SECOs. The mapping study revealed that the research on SECOs has progressed well since 2009, from then at least four publications were focusing on RE in SECOs each year, and this is a potential sign that the research on RE of SECOs is gaining increasing attention among researchers and practitioners in software ecosystems.

Regarding to RQ1, our findings indicate that the most addressed topics on RE in SECO are identifying stakeholders and roles for requirements elicitation and requirements analysis. Especially, half of the collected papers from 2011 tended to address requirements negotiation. However, the requirements selection during the integration of components, prioritization were not studied in the existing literature.

Regarding to RQ2 our findings indicate that the most addressed topics on non-functional requirements are related to security, performance and testability (published more than 2 papers). We found that there are two types of quality concerns: In-operation and In-development. The most addressed topics on the quality attributes security, performance and testability related to in-operation are certification and policy management security patterns, software redundancy, authentication, accountability, transparency. The most addressed topics on the quality attributes security, performance and testability related to in-development are developer performance, performance measurements of software ecosystems, ecosystem healthiness, and software testing requirements, static analysis of vulnerabilities in web applications and their plug-ins, review and approval methods for platform extensions.

For practitioners (platform providers or suppliers, extenders i.e. developers and users), who want to explore quality aspects in SECOs should focus more on documentation of platform features and architecture, dependency and maintainability mechanisms, guidelines and policies of platform adoption, taking decisions on the level of transparency to enhance the design, and usability issues.

For researchers who want to break new ground in quality aspects of platform extensions in SECOs, a possible focus could be on quality attributes such as reliability, maintainability, transparency and usability, where this mapping study indicated that there has been little or no research so far. Moreover, even for the areas having most publications within RE for SECOs, the number of publications found in this mapping study is not immense compared to the interest and industrial relevance of the topic. Especially, the amount of publications with heavy empirical work is limited so far, so any such study would be welcome, also related to the topics which appeared to have better than average coverage in the mapping study reported here.

**Appendix A. Literature body**

1.  Bosch and Bosch-Sijtsema (2010)
2.  Bosch (2009)
3.  Jansen (2013)
4.  Manikas and Hansen (2013)
5.  Bosch-Sijtsema and Bosch (2015)
6.  Hull et al. (2011)
7.  Serebrenik and Mens (2015)
8.  Barbosa and Alves (2013)
9.  Jansen et al. (2009)
10. Jansen et al. (2009)
11. Manikas (2016)
12. Scacchi and Alspaugh (2012)
13. Bosch (2010)
14. Hartmann and Bosch (2014)
15. García-Holgado and García-Peñalvo (2016)
16. Rajeshwar (2017)
17. Fricker (2010)
18. Van den Berk et al. (2010)
19. Valença et al. (2014)
20. Angeren et al. (2011)
21. Cataldo and Herbsleb (2010)
22. Knauss et al. (2014)
23. Goeminne (2014)
24. Santos and Werner (2012)
25. Scacchi (2007)
26. Teixeira and Lin (2014)
27. Hanssen and Dybå (2012)
28. Loucopoulos and Karakostas (1995)

29. Shewhart (1930)
30. Pressman (2005)
31. McCall et al. (1977)
32. Boehm et al. (1978)
33. Dromey (1995)
34. Standardization and Commission (2001)
35. Kitchenham et al. (2011)
36. Nguyen-Duc (2017)
37. Nguyen-Duc et al. (2015)
38. Axelsson and Skoglund (2016)
39. Garousi et al. (2016)
40. Gherbi et al. (2011)
41. Eklund and Bosch (2014)
42. Mhamdia (2013)
43. Christensen et al. (2014)
44. Fernandez et al. (2016)
45. Leite and Cappelli (2010)
46. Santos (2014)
47. Goldbach and Benlian (2015)
48. Fernandez et al. (2015)
49. Pettersson et al. (2010)
50. Sadi and Yu (2017)
51. H. Sadi and Yu (2015)
52. Handoyo et al. (2013)
53. Scacchi and Alspaugh (2012)
54. Schultis et al. (2014)
55. Bezzi et al. (2013)
56. Claes et al. (2014)
57. Fahl et al. (2014)
58. Hansen and Zhang (2013)
59. Dantas et al. (2009)
60. Valenca (2013)
61. Campbell and Ahmed (2010)
62. Knauss et al. (2016)
63. Jansen et al. (2015)
64. Soltani and Knauss (2015)
65. Fricker (2009)
66. Angeren et al. (2011)
67. Yu and Deng (2011)
68. Jansen (2013)
69. Schultis et al. (2013)
70. Boucharas et al. (2009)
71. Walden et al. (2010)
72. Jansen and van Capelleveen (2013)
73. Lettner and Grünbacher (2015)

**References**

Angeren, J. v., Blijleven, V., & Jansen, S. (2011). *Relationship intimacy in software ecosystems: a survey of the Dutch software industry*. Paper presented at the Proceedings of the International Conference on Management of Emergent Digital EcoSystems, San Francisco, California.

Angeren, J. v., Kabbedijk, J., Jansen, S., & Popp, K. M. (2011). *A survey of associate models used within large software ecosystems*. Paper presented at the Proceedings of the Third International Workshop on Software Ecosystems(IWSECO2011), , Brussels, Belgium.

Axelsson, J., & Skoglund, M. (2016). Quality assurance in software ecosystems: A systematic literature mapping and research agenda. *Journal of Systems and Software, 114*, 69-81.

Barbosa, O., & Alves, C. (2013). A Systematic Mapping Study on Software Ecosystems through a Three-dimensional Perspective (pp. 59-81). Software Ecosystems: Analyzing and Managing Business Networks in the Software Industry.

Bezzi, M., Damiani, E., Paraboschi, S., & Plate, H. (2013, 2013). *Integrating Advanced Security Certification and Policy Management*. Paper presented at the Felici M. (eds) Cyber Security and Privacy. CSP

Boehm, B., Brown, J., Kaspar, H., Lipow, M., MacLeod, G., & Merritt, M. (1978). *Characteristics of Software Quality*: North-Holland.

Bosch-Sijtsema, P., & Bosch, J. (2015). Plays nice with others? Multiple ecosystems, various roles and divergent engagement models. *Techn. Analysis & Strat. Manag., 27*(8), 960-974.

Bosch, J. (2009). *From software product lines to software ecosystems*. Paper presented at the Proceedings of the 13th International Software Product Line Conference, San Francisco, California, USA.

Bosch, J. (2010). *Architecture challenges for software ecosystems*. Paper presented at the Proceedings of the Fourth European Conference on Software Architecture: Companion Volume, Copenhagen, Denmark.

Bosch, J., & Bosch-Sijtsema, P. (2010). From integration to composition: On the impact of software product lines, global development and ecosystems. *Journal of Systems and Software, 83*(1), 67-76.

Boucharas, V., Jansen, S., & Brinkkemper, S. (2009). *Formalizing software ecosystem modeling*. Paper presented at the Proceedings of the 1st international workshop on Open component ecosystems, Amsterdam, The Netherlands.

Campbell, P. R. J., & Ahmed, F. (2010). *A Three-dimensional View of Software Ecosystems*. Paper presented at the Proceedings of the Fourth European Conference on Software Architecture.

Cataldo, M., & Herbsleb, J. D. (2010). *Architecting in Software Ecosystems: Interface Translucence As an Enabler for Scalable Collaboration*.

Christensen, H. B., Hansen, K. M., Kyng, M., & Manikas, K. (2014). Analysis and design of software ecosystem architectures – Towards the 4S telemedicine ecosystem. *Information and Software Technology, 56*(11), 1476-1492.

Claes, M., Mens, T., & Grosjean, P. (2014). *On the maintainability of CRAN packages*. Paper presented at the 2014 Software Evolution Week - IEEE Conference on Software Maintenance, Reengineering, and Reverse Engineering, CSMR-WCRE 2014 - Proceedings.

Dantas, V. L. L., Marinho, F. G., Costa, A. L. d., & Andrade, R. M. C. (2009). *Testing requirements for mobile applications*. Paper presented at the 2009 24th International Symposium on Computer and Information Sciences.

Dromey, R. G. (1995). A Model for Software Product Quality. *IEEE Trans. Softw. Eng., 21*(2), 146-162.

Eklund, U., & Bosch, J. (2014). Architecture for embedded open software ecosystems. *Journal of Systems and Software, 92*, 128-142.

Fahl, S., Dechand, S., Perl, H., Fischer, F., Smrcek, J., & Smith, M. (2014). *Hey, NSA: Stay Away from My Market! Future Proofing App Markets Against Powerful Attackers*. Paper presented at the Proceedings of the 2014 ACM SIGSAC Conference on Computer and Communications Security.

Fernandez, E. B., Yoshioka, N., & Washizaki, H. (2015). *Patterns for security and privacy in cloud ecosystems*. Paper presented at the Evolving Security and Privacy Requirements Engineering (ESPRE), 2015 IEEE 2nd Workshop on.

Fernandez, E. B., Yoshioka, N., Washizaki, H., & Syed, M. H. (2016). Modeling and security in cloud ecosystems. *Future Internet, 8*(2).

Fricker, S. (2009). *Specification and Analysis of Requirements Negotiation Strategy in Software Ecosystems.* Paper presented at the in IWSECO@ ICSR.

Fricker, S. (2010). Requirements Value Chains: Stakeholder Management and Requirements Engineering in Software Ecosystems. In R. Wieringa & A. Persson (Eds.), *Requirements Engineering: Foundation for Software Quality* (pp. 60-66): Springer Berlin Heidelberg.

García-Holgado, A., & García-Peñalvo, F. J. (2016). Architectural pattern to improve the definition and implementation of eLearning ecosystems. *Science of Computer Programming, 129*, 20-34.

Garousi, V., Felderer, M, M, M. V., #228, ntyl, & #228. (2016). *The need for multivocal literature reviews in software engineering: complementing systematic literature reviews with grey literature.* Paper presented at the Proceedings of the 20th International Conference on Evaluation and Assessment in Software Engineering, Limerick, Ireland.

Gherbi, A., Charpentier, R., & Couture, M. (2011). Software diversity for future systems security. *CrossTalk, 24*(5), 10-13.

Goeminne, M. (2014). *Understanding the evolution of socio-technical aspects in open source ecosystems.* Paper presented at the 2014 Software Evolution Week - IEEE Conference on Software Maintenance, Reengineering, and Reverse Engineering, CSMR-WCRE 2014 - Proceedings.

Goldbach, T., & Benlian, A. (2015). *Understanding informal control modes on software platforms -The mediating role of third-party developers' intrinsic motivation.* Paper presented at the 2015 International Conference on Information Systems: Exploring the Information Frontier, ICIS 2015.

H. Sadi, M., & Yu, E. (2015). Designing Software Ecosystems: How Can Modeling Techniques Help? In K. Gaaloul, R. Schmidt, S. Nurcan, S. Guerreiro, & Q. Ma (Eds.), *Enterprise, Business-Process and Information Systems Modeling: 16th International Conference, BPMDS 2015, 20th International Conference, EMMSAD 2015, Held at CAiSE 2015, Stockholm, Sweden, June 8-9, 2015, Proceedings* (pp. 360-375). Cham: Springer International Publishing.

Handoyo, E., Jansen, S., & Brinkkemper, S. (2013) Software ecosystem roles classification. *Vol. 150 LNBIP. Lecture Notes in Business Information Processing* (pp. 212-216).

Hansen, K. M., & Zhang, W. (2013). *Towards structure-based quality awareness in software ecosystem use.* Paper presented at the International Conference on Service-Oriented Computing, ICSOC, Berlin, Germany.

Hanssen, G. K., & Dybå, T. (2012). *Theoretical foundations of software ecosystems.* Paper presented at the In Proceedings of IWSECO, Boston.

Hartmann, H., & Bosch, J. (2014). Orchestrate Your Platform: Architectural Challenges for Different Types of Ecosystems for Mobile Devices. In C. Lassenius & K. Smolander (Eds.), *Software Business. Towards Continuous Value Delivery* (pp. 163-178): Springer International Publishing.

Hull, E., Jackson, K., & Dick, J. (2011). *Requirements Engineering.* London: Springer London.

Jansen, S. (2013). *How quality attributes of software platform architectures influence software ecosystems.* Paper presented at the Proceedings of the 2013 International Workshop on Ecosystem Architectures, Saint Petersburg, Russia.

Jansen, S., and Michael A. Cusumano. (2013). Defining software ecosystems: a survey of software platforms and business network governance. *Software Ecosystems: Analyzing and Managing Business Networks in the Software Industry 13.*

Jansen, S., Brinkkemper, S., & Finkelstein, A. (2009). *Business Network Management as a Survival Strategy: A Tale of Two Software Ecosystems.* Paper presented at the First International Workshop on Software Ecosystems (IWSECO).

Jansen, S., Finkelstein, A., & Brinkkemper, S. (2009). *A sense of community: A research agenda for software ecosystems.* Paper presented at the 31st International Conference on Software Engineering - Companion Volume, 2009. ICSE-Companion 2009.

Jansen, S., Handoyo, E., & Alves, C. (2015). *Scientists' Needs in Modelling Software Ecosystems.* Paper presented at the Proceedings of the 2015 European Conference on Software Architecture Workshops, Dubrovnik, Cavtat, Croatia.

Jansen, S., & van Capelleveen, G. (2013). Quality review and approval methods for extensions in software ecosystems *Software Ecosystems: Analyzing and Managing Business Networks in the Software Industry* (pp. 187-217): Edward Elgar Publishing.

Kitchenham, B. A., Budgen, D., & Pearl Brereton, O. (2011). Using mapping studies as the basis for further research – A participant-observer case study. *Information and Software Technology, 53*(6), 638-651.

Knauss, E., Damian, D., Knauss, A., & Borici, A. (2014). *Openness and requirements: Opportunities and tradeoffs in software ecosystems.* Paper presented at the 2014 IEEE 22nd International Requirements Engineering Conference (RE).

Knauss, E., Yussuf, A., Blincoe, K., Damian, D., & Knauss, A. (2016). Continuous clarification and emergent requirements flows in open-commercial software ecosystems. *Requirements Engineering*, 1-21.

Leite, J. C. S. d. P., & Cappelli, C. (2010). Software Transparency. *Business & Information Systems Engineering, 2*(3), 127-139.

Lettner, D., & Grünbacher, P. (2015). *Using Feature Feeds to Improve Developer Awareness in Software Ecosystem Evolution.* Paper presented at the Proceedings of the Ninth International Workshop on Variability Modelling of Software-intensive Systems, Hildesheim, Germany.

Loucopoulos, P., & Karakostas, V. (1995). *Systems Requirements Engineering*: McGraw-Hill.

Manikas, K. (2016). Revisiting software ecosystems Research: A longitudinal literature study. *Journal of Systems and Software, 117*, 84-103.

Manikas, K., & Hansen, K. M. (2013). Software ecosystems – A systematic literature review. *Journal of Systems and Software, 86*(5), 1294-1306.

McCall, J. A., Richards, P. K., & Walters, G. F. (1977). *Factors in software quality. volume i. concepts and definitions of software quality*.

Mhamdia, A. B. H. S. (2013). Performance measurement practices in software ecosystem. *International Journal of Productivity and Performance Management, 62*(5), 514-533.

Nguyen-Duc, A. (2017). The Impact of Software Complexity on Cost and Quality - A Comparative Analysis Between Open Source and Proprietary Software. *International Journal on Software Engineering and Application, 8*(2), 17-31.

Nguyen-Duc, A., Cruzes, D. S., & Conradi, R. (2015). The impact of global dispersion on coordination, team performance and software quality – A systematic literature review. *Information and Software Technology, 57*, 277-294.

Pettersson, O., Svensson, M., Gil, D., Andersson, J., & Milrad, M. (2010). *On the Role of Software Process Modeling in Software Ecosystem Design.* Paper presented at the Proceedings of the Fourth European Conference on Software Architecture: Companion Volume.

Pressman, R. S. (2005). *Software engineering: a practitioner's approach*: Palgrave Macmillan.

Rajeshwar, V. (2017). Software Engineering for Technological Ecosystems. In J. G.-P. Francisco & G.-H. Alicia (Eds.), *Open Source Solutions for Knowledge Management and Technological Ecosystems* (pp. 175-194). Hershey, PA, USA: IGI Global.

Sadi, M. H., & Yu, E. (2017). *Modeling and analyzing openness trade-offs in software platforms: A goal-oriented approach*. Paper presented at the Grünbacher P., Perini A. (eds) Requirements Engineering: Foundation for Software Quality, REFSQ.

Santos, R. P. d. (2014). *ReuseSEEM: an approach to support the definition, modeling, and analysis of software ecosystems*. Paper presented at the Companion Proceedings of the 36th International Conference on Software Engineering, Hyderabad, India.

Santos, R. P. d., & Werner, C. M. L. (2012). *ReuseECOS: An Approach to Support Global Software Development through Software Ecosystems.* Paper presented at the 2012 IEEE Seventh International Conference on Global Software Engineering Workshops (ICGSEW).

Scacchi, W. (2007). *Free/open source software development: Recent research results and emerging opportunities.* Paper presented at the Proceedings of the the 6th joint meeting of the European Software Engineering Conference and the ACM SIGSOFT Symposium on the Foundations of Software Engineering 2007, ESEC-FSE'07.

Scacchi, W., & Alspaugh, T. A. (2012a, 2012/09/10/). *Designing Secure Systems Based on Open Architectures with Open Source and Closed Source Components.* Paper presented at the International Conference on Open Source Systems.

Scacchi, W., & Alspaugh, T. A. (2012b). Understanding the role of licenses and evolution in open architecture software ecosystems. *Journal of Systems and Software, 85*(7), 1479-1494.

Schultis, K.-B., Elsner, C., & Lohmann, D. (2014). *Architecture Challenges for Internal Software Ecosystems: A Large-scale Industry Case Study.* Paper presented at the Proceedings of the 22nd ACM SIGSOFT International Symposium on Foundations of Software Engineering.

Schultis, K. B., Elsner, C., & Lohmann, D. (2013). *Moving towards industrial software ecosystems: Are our software architectures fit for the future?* Paper presented at the 2013 4th International Workshop on Product Line Approaches in Software Engineering (PLEASE).

Serebrenik, A., & Mens, T. (2015). *Challenges in Software Ecosystems Research.* Paper presented at the Proceedings of the 2015 European Conference on Software Architecture Workshops, Dubrovnik, Cavtat, Croatia.

Shewhart, W. A. (1930). Economic quality control of manufactured product. *The Bell System Technical Journal, 9*(2), 364-389.

Soltani, M., & Knauss, E. (2015). *Cross-organizational challenges of requirements engineering in the AUTOSAR Ecosystem: An exploratory case study.* Paper presented at the 2015 IEEE Fifth International Workshop on Empirical Requirements Engineering (EmpiRE).

Standardization, I. O. f., & Commission, I. E. (2001). *Software Engineering--Product Quality: Quality model* (Vol. 1): ISO/IEC.

Teixeira, J., & Lin, T. (2014). *Collaboration in the Open-source Arena: The Webkit Case*.

Valenca, G. (2013). *Requirements negotiation model: A social oriented approach for software ecosystems evolution.* Paper presented at the Requirements Engineering Conference (RE), 2013 21st IEEE International.

Valença, G., Carina, A., Virgínia, H., Slinger, J., & Sjaak, B. (2014). *Competition and collaboration in requirements engineering: A case study of an emerging software ecosystem.* Paper presented at the Requirements Engineering Conference (RE), 2014 IEEE 22nd International.

Van den Berk, I., Jansen, S., & Luinenburg, L. (2010). *Software ecosystems: a software ecosystem strategy assessment model.* Paper presented at the Proceedings of the Fourth European Conference on Software Architecture: Companion Volume, Copenhagen, Denmark.

Walden, J., Doyle, M., Lenhof, R., Murray, J., & Plunkett, A. (2010). *Impact of Plugins on the Security of Web Applications.* Paper presented at the Proceedings of the 6th International Workshop on Security Measurements and Metrics, Bolzano, Italy.

Yu, E., & Deng, S. (2011). Understanding software ecosystems: A strategic modeling approach. *proc of 3rd IWSECO*, 65-76.